\documentclass[doublespacing]{elsart}

\usepackage{graphicx}
\usepackage{amssymb}

\begin{document}
\begin{frontmatter}

\journal{SCES'2001: Version 1}


\title{Anisotropic strains, metal-insulator transition, and
magnetoresistance of La$_{0.7}$Ca$_{0.3}$MnO$_{3}$ films}

\author[essc]{J. H. Song} \author[essc]{J. -H. Park}
 \author[essc]{Y. H. Jeong\corauthref{1}}
 \author[rutgers]{T. Y. Koo}

 \address[essc]{Department of Physics and electron Spin Science Center,
 POSTECH\thanksref{bk}}
 \address[rutgers]{Department of Physics, Rutgers University}

\thanks[bk]{Work at POSTECH was supported by the SRC program of KOSEF and the BK21 program of MOE.}

\corauth[1]{Corresponding Author: electron Spin Science Center,
 Pohang Univ. of Sci and Tech, Pohang 790-784, S.
 Korea,  Fax: +82-54-279-8056, Email: yhj@postech.ac.kr}


\begin{abstract}
Thin films of perovskite manganite La$_{0.7}$Ca$_{0.3}$MnO$_{3}$
were grown epitaxially on various substrates by either the pulsed
laser deposition method or laser molecular beam epitaxy. The
substrates change both the volume and symmetry of the unit cell of
the films. It is revealed that the symmetry as well as the volume
of the unit cell have strong influence on the metal-insulator
transition temperature and the size of magnetoresistance.
\end{abstract}


\begin{keyword}
 anisotropic strains\sep metal-insulator transition \sep
magnetoresistance

\end{keyword}


\end{frontmatter}


 The perovskite compound La$_{1-x}$Ca$_{x}$MnO$_3$ with $x\approx 0.3$ (LCMO) shows a simultaneous
appearance of metallic conduction and ferromagnetism below a
certain ordering temperature \cite{jonker,wollan}. It has
attracted renewed attention due to its anomalously large
magnetoresistance (MR) \cite{jin1}. Considering numerous potential
applications of large MR materials in thin film form, it would be
of value to carefully examine their properties as a function of
lattice strain \cite{koo}.

Here we briefly describe our systematic investigations on the
strain effects on $T_{\rm p}$, the temperature at which a maximum
resistance at zero field occurs, and MR of LCMO. By exploiting the
fact that the lattice parameters of films can be varied by
depositing films on various substrates, we were able to vary the
lattice symmetry and constants of LCMO. Two methods of film
preparation were used, pulsed laser deposition (PLD) and laser
molecular beam epitaxy (LMBE). PLD and LMBE are similar in that
high power pulsed laser is used as an energy source. However,
there exist crucial differences in growth rate and growth mode; in
PLD the typical growth rate is approximately 3 ${\rm \AA/s}$ and
films grow in the 3 dimensional island mode, while the growth rate
of LMBE is about 0.1 ${\rm \AA/s}$ and films grow in the
layer-by-layer mode. The layer-by-layer mode for LMBE is
ascertained by in-situ monitoring with reflective high energy
electron diffraction; the details of the LMBE method will be
described elsewhere \cite{song}.

A set of LCMO films with thickness 1000 ${\rm \AA}$ were grown by
PLD on SrTiO$_3$(100), MgO(100), and LaAlO$_3$(100) substrates
(designated as the PLD films), and another set of thin films with
thickness 400 ${\rm \AA}$ were deposited by LMBE on
SrTiO$_3$(100), NdGaO$_3$(100), and LaAlO$_3$(100) substrates
(LMBE films). We first turn our attention to the results of four
probe resistance measurements. In Fig.~1(a) plotted is the
resistance $R$ of the the PLD films at $B$ = 0 and 1 T  versus
$T$; Fig.~1(b) is the plot of the resistance of the LBME films. It
is striking that $T_{\rm p}$, the maximum resistance temperature,
varies enormously even within the set of the films prepared by the
same method. This indicates that the strains in the films on
various substrates vary and thus cause a variation in $T_{\rm p}$
despite the fact that the processing conditions are exactly the
same within a given set of films. High resolution 4-circle x-ray
diffraction was carried out at room temperature to determine the
strains, and the in-plane and surface-normal lattice parameters of
the LCMO films are designated as $a$ and $c$, respectively. It
should be noted here that the lattice parameters of ultra-thin
films such as those under discussion may vary as a function of
distance from the substrate \cite{zhu}, and if this is the case,
then the lattice parameters measured may be regarded as the
averaged ones.

For the PLD films, it is found that the film on SrTiO$_3$
possesses the largest unit cell volume ($V_{\rm c}$) and the one
on LaAlO$_3$ does the smallest. In addition, the lattice symmetry
of the films changes from negative tetragonal (the SrTiO$_3$ case,
$a$=3.921 ${\rm \AA}$, $c$=3.845 ${\rm \AA}$) to nearly cubic (the
MgO case, $a$=3.885 ${\rm \AA}$, $c$=3.891 ${\rm \AA}$) and to
positive or elongated tetragonal one (the LaAlO$_3$ case,
$a$=3.842 ${\rm \AA}$, $c$=3.878 ${\rm \AA}$) as unit cell volume
decreases. To expose the correlation between the lattice structure
and $T_{\rm p}$ and MR of LCMO, we plot $T_{\rm p}$ and the
maximum MR values against unit cell volume. Fig.~2(a) clearly
illustrates that the samples with cubic symmetry (open symbols)
show lower values of maximum MR than the noncubic ones (filled
symbols) do. Fig.~2(b) shows that $T_{\rm p}$ also varies
nonmonotonically. This distinctive $T_{p}$ behavior may be
interpreted as a superposition of two differing tendencies: one
tendency is related to the {\em volume effect} favoring the
monotonic increase of $T_{\rm p}$ as the volume of unit cell is
reduced. The other tendency is due to the {\em symmetry effect},
in accord with the MR behavior, favoring the maximum $T_{\rm p}$
for cubic samples.

It is noted that the volume effect resembles that of a bulk under
a hydrostatic pressure where the reduction in unit cell volume
gives rise to a monotonic $T_{\rm p}$ rise and an associated
decrease of maximum MR \cite{hwang2,khazeni}. However, the
symmetry effect cannot be probed with isotropic hydrostatic
pressure. To further shed light on the symmetry effect, we
examined the LMBE films, i.e., the films grown in the
layer-by-layer mode on SrTiO$_3$, NdGaO$_3$, and LaAlO$_3$
substrates. From the x-ray diffraction measurements, it was found
that the unit cell volume variation in the LMBE films is within
0.8$\%$, about four times less than that of the PLD films. X-ray
diffraction also showed that the lattice symmetry of the films
again changes from negative tetragonal (the SrTiO$_3$ case,
$a$=3.893 ${\rm \AA}$, $c$=3.827 ${\rm \AA}$) to nearly cubic (the
NdGaO$_3$ case, $a$=3.865 ${\rm \AA}$, $c$=3.874 ${\rm \AA}$) and
to positive tetragonal one (the LaAlO$_3$ case, $a$=3.838 ${\rm
\AA}$, $c$=3.903 ${\rm \AA}$). Fig.~3 is the plot of $T_{\rm p}$
against the teragonality. The figure clearly reveals the symmetry
effect; a deviation from the cubic symmetry leads to a reduction
in the transition temperature. It is of value to point out that
the negative tetragonality is more effective in reducing $T_{\rm
p}$ than the positive one. This, we believe, is related to the Mn
d-orbital ordering induced by the imposed strains, and will be the
topic of future publications.




\begin{figure}
\centering
\includegraphics[height=14cm,width=10cm]{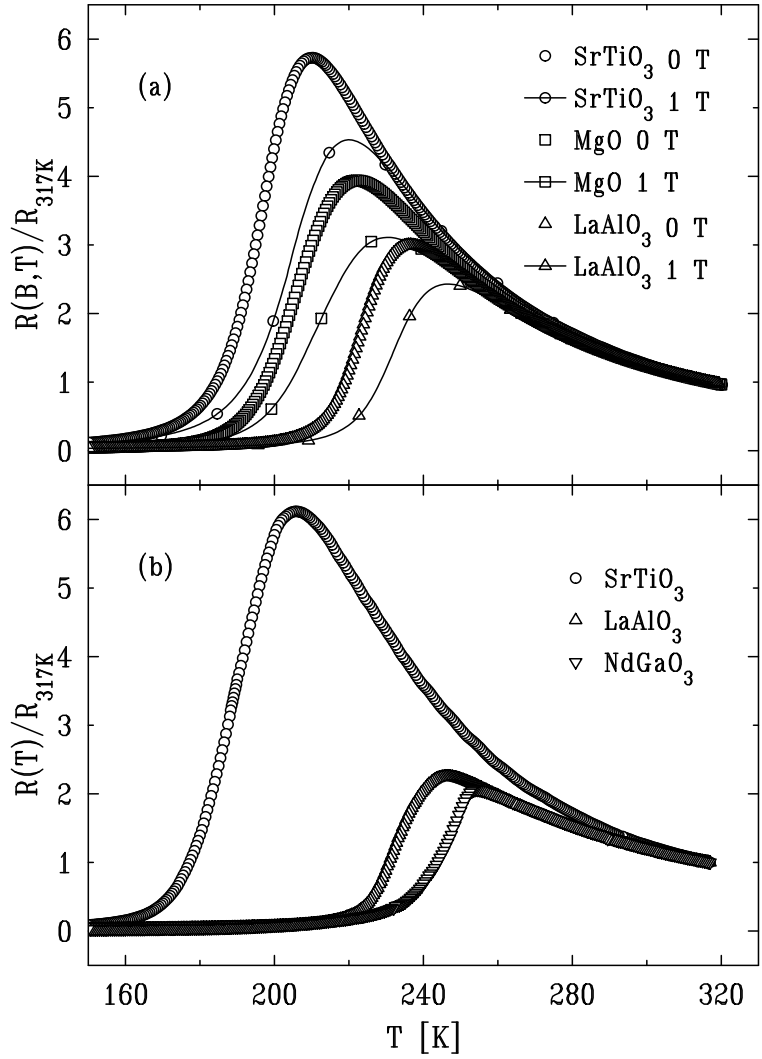}
\caption{Plotted as a function of temperature ($T$) is resistivity
of La$_{0.7}$Ca$_{0.3}$MnO$_3$ thin films: (a) films deposited on
SrTiO$_3$, MgO, and LaAlO$_3$ by the PLD method, and (b) films
deposited on SrTiO$_3$, NdGaO$_3$, and LaAlO$_3$ by LMBE.
Resistivity of each sample is normalized to the value at $T$= 317
K and zero field.}
\end{figure}

\begin{figure}
\centering
\includegraphics[height=14cm,width=10cm]{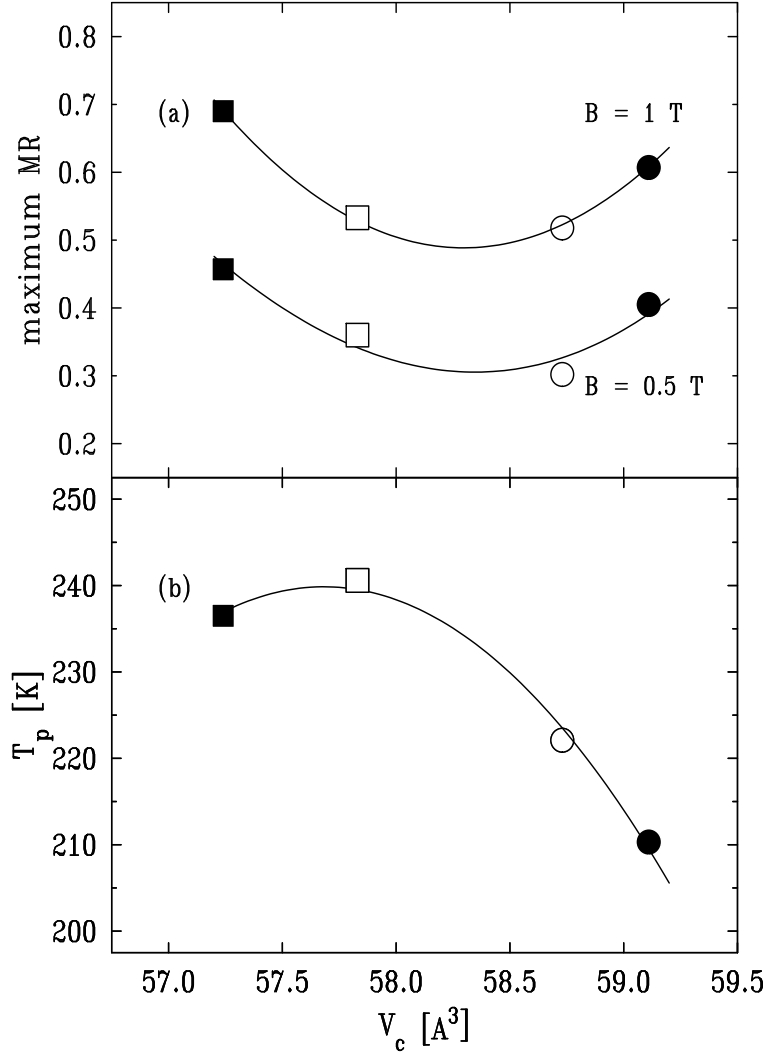}
\caption{MR ($\equiv [R(0) - R(B)]/R(0)$) and $T_{\rm p}$ of the
films grown by the PLD method on SrTiO$_3$ (filled circle), MgO
(open circle), and LaAlO$_3$ (filled square). The open square
denotes the bulk values. (a) The maximum MR at $B$ = 0.5 and 1 T,
and (b) $T_{\rm p}$ at zero field are plotted against the unit
cell volume V$_c$. The lines are guide for the eye.}
\end{figure}

\begin{figure}
\centering
\includegraphics[width=10cm]{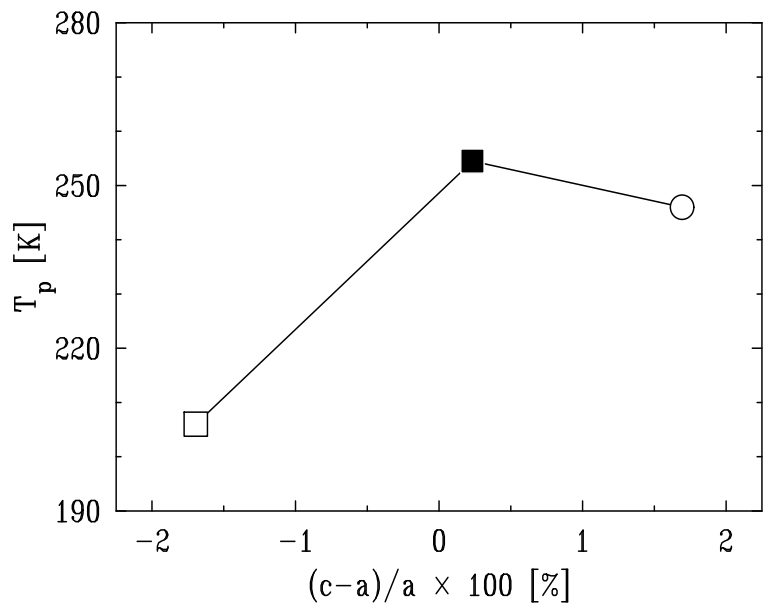}
\caption{$T_{\rm p}$ of the films grown by LMBE is plotted as a
function of tetragonality $(c-a)/a$ where $c$ and $a$ indicate the
normal and in-plane lattice constants, respectively. Different
symbols indicates the substrates: SrTiO$_3$ (open square),
NdGaO$_3$ (filled square), and LaAlO$_3$ (open circle). The line
is guide for the eye.}
\end{figure}


\end{document}